\documentclass[twocolumn]{aastex62}
\usepackage{amsmath}

\newcommand{\rtwo}{$r_{20}$}
\newcommand{\reig}{$r_{80}$}
\newcommand{\rfiv}{$r_{\rm 50}$}

\graphicspath{{./}{figures/}}

\received{XXX}
\revised{YYY}
\accepted{ZZZ}
\submitjournal{ApJL}

\shorttitle{The distribution of \rtwo\ and \reig}
\shortauthors{Miller et al.}

\begin{document}

\title{A New View of the Size-Mass Distribution of Galaxies: Using $r_{20}$ and $r_{80}$ instead of $r_{50}$}

\correspondingauthor{Tim B. Miller}
\email{tim.miller@yale.edu}

\author{Tim B. Miller}
\affil{Department of Astronomy, Yale University, 52 Hillhouse Ave., New Haven, CT, USA, 06511 }

\author{Pieter van Dokkum}
\affil{Department of Astronomy, Yale University, 52 Hillhouse Ave., New Haven, CT, USA, 06511 }

\author{Lamiya Mowla}
\affil{Department of Astronomy, Yale University, 52 Hillhouse Ave., New Haven, CT, USA, 06511 }

\author{Arjen van der Wel}
\affil{Sterrenkundig Observatorium,  Universiteit Gent,  Krijgslaan 281 S9,B-9000 Gent, Belgium}
\affil{Max-Planck-Institut f{\"u}r Astronomie, K{\"o}nigstuhl 17, D-69117, Heidelberg, Germany}

\begin{abstract}

When investigating the sizes of galaxies it is standard practice to use the half-light radius, $r_{50}$. Here we explore the effects of the size definition on the distribution of galaxies in the size -- stellar mass plane. Specifically, we consider $r_{20}$ and $r_{80}$, the radii that contain 20\% and 80\% of a galaxy's total luminosity, as determined from a Sersic profile fit, for galaxies in the 3D-HST/CANDELS and COSMOS-DASH surveys. These radii are calculated from size catalogs based on a simple calculation assuming a Sersic profile. We find that the size-mass distributions for $r_{20}$ and $r_{80}$ are markedly different from each other and also from the canonical $r_{50}$ distribution. The most striking difference is in the relative sizes of star forming and quiescent galaxies at fixed stellar mass. Whereas quiescent galaxies are smaller than star forming galaxies in $r_{50}$, this difference nearly vanishes for $r_{80}$.  By contrast, the distance between the two populations {\em increases} for $r_{20}$. 
Considering all galaxies in a given stellar mass and redshift bin we detect a significant bimodality in the distribution of $r_{20}$, with one peak corresponding to star forming galaxies and the other to quiescent galaxies.
We suggest that different measures of the size are tracing different physical processes within galaxies; $r_{20}$ is closely related to processes controlling the star formation rate of galaxies and $r_{80}$ may be sensitive to accretion processes and the relation of galaxies with their halos.
\end{abstract}

\keywords{galaxies: Structure --- galaxies: fundamental parameters --- galaxies: high-redshift}

\section{Introduction} \label{sec:intro}

The sizes of galaxies hold clues about the physical processes which shape them. They can be predicted by galaxy formation models~\citep{Mo1998,Dutton2011,Kravtsov2013,Somerville2018} and can help distinguish between different evolutionary models~\citep{Carollo2013,vandokkum2015, Matharu2018}. 
However, the sizes of galaxies are difficult to define, as their surface brightness profiles decrease smoothly with radius with no well-defined edge. A common method is to use the half-light, also known as the effective radius, \rfiv, which contains 50\% of a galaxy's total luminosity. It is generally applicable to all galaxies and does not trivially correlate with other properties such as a galaxy's luminosity. Due to these properties, \rfiv\ has become the standard measurement of the size of a galaxy. Studies of \rfiv\ over the past decades have shown that it correlates with stellar mass, the so called size-mass distribution, which in turn varies with galaxy colour, type and redshift~\citep{Shen2003,Ferguson2003,Trujillo2006,Williams2010,Ono2013,vanderwel2014,Lange2015,Mowla2018}. 

When investigating the size-mass distribution it is important to assess the effect of the choice of the size parameter, as a single number fails to capture information about the distribution of light within a galaxy. In practice, a second parameter is typically introduced to separately study the form of the light profile. The Sersic index $n$ \citep{Sersic1968} has become the parameter of choice, derived from one- or two-dimensional fits of the form $\log I(r) \propto (r/r_{50})^{1/n}$ to the surface brightness profile.

In this study we explore an alternative approach to studying the structure of galaxies. We compare and contrast the size-mass distribution that arises from using different measures for the size of a galaxy. We will use \rtwo\ and \reig, the radii that contain 20\% and 80\% of the total luminosity, along with the canonical measure of \rfiv. This study will focus on the difference between star-forming and quiescent galaxies at a fixed stellar mass to investigate the different evolutionary processes which shape them. In a accompanying paper, Mowla et al.\ (2019),
we investigate the relation between \reig\ and a galaxy's dark matter halo.

\section{Data}
\subsection{Galaxy Sample} \label{sec:Sample}

In this study we employ two different galaxy surveys: 3D-HST/CANDELS~\citep{Koekemoer2011,Brammer2012} and COSMOS-DASH~\citep{Momcheva2016,Mowla2018}. The CANDELS survey covers 0.22 degree$^2$ with extensive ground and space based photometry ranging from $0.3\,\mu m-8\,\mu m$, which is supplemented by WFC3 grism spectroscopy spanning three quarters of that area. Galaxy sizes are measured in \citet{vanderwel2014} from the $H_{160}$ and $I_{814}$ bands for $\sim 30,000$ galaxies above $M_* > 10^9 M_\odot$ with $0<z<3$. Galaxy properties such as stellar mass, redshift and rest-frame colors for this sample are taken from the 3D-HST catalog~\citep{Skelton2014}. We supplement this sample with the COSMOS-DASH survey which covers 0.66 deg$^2$ with $H_{160}$ imaging. The larger survey area affords proper sampling of the bright end of the luminosity function for $1.5<z<3$ which is not possible in the smaller CANDELS survey. Combined with 1.7 deg$^2$ of ACS-COSMOS imaging~\citep{Koekemoer2007}, \citet{Mowla2018} measure the sizes of $910$ galaxies with $M_* > 2\times 10^{11} M_\odot$ at $0<z<3$. Masses and redshifts for the COSMOS-DASH sample are taken from the UltraVISTA catalog~\citep{Muzzin2013a}, as described in \citet{Mowla2018}.

\citet{vanderwel2014} and \citet{Mowla2018} use very similar methods to measure the size of galaxies. \texttt{GALFIT}~\citep{Peng2010} is used to fit two-dimensional single component Sersic profiles to each galaxy and extract a best fit Sersic index and effective radius. This forward modelling approach allows the measurement of galaxy sizes which are comparable to the instrumental point spread function (PSF). The ACS/F814W filter is used for galaxies with $z<1.5$ and the WFC3/F160W filter is used at higher redshift. Redshift- and mass-dependent color gradients are taken into account to ensure that the sizes of all galaxies are measured at the same rest-frame wavelength ($5000$\AA). Throughout this study we will separate galaxies into two populations: star-forming and quiescent. This is done using their rest-frame UVJ colours according to the prescription in \citet{Muzzin2013}.

\subsection{Calculating \rtwo\ and \reig}
\begin{figure}
    \centering
    \includegraphics[width = \columnwidth]{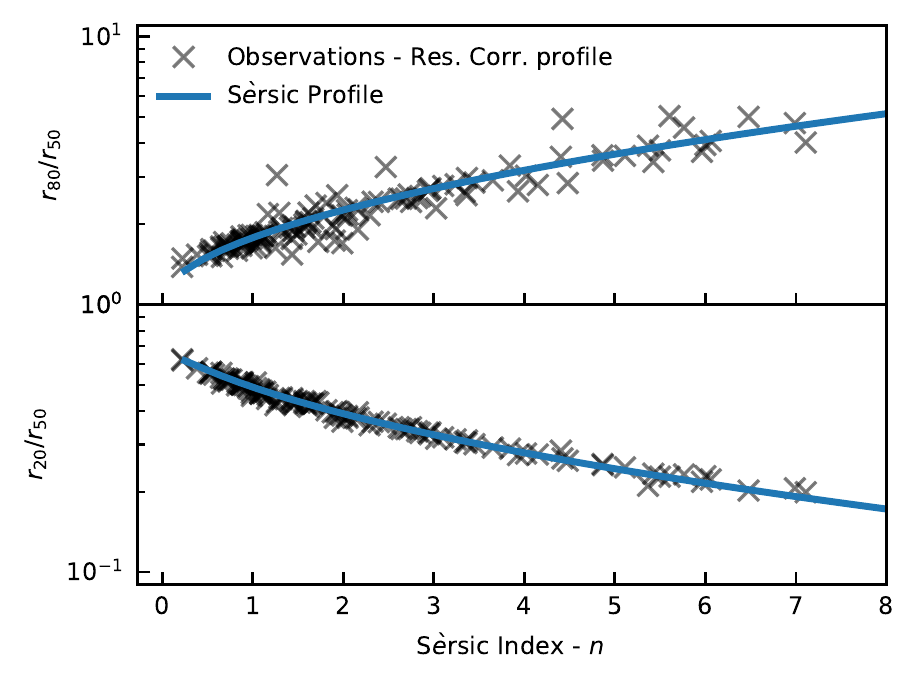}
    \caption{The ratios $r_{20\%}/r_{\rm eff}$ and $r_{80\%}/r_{\rm eff}$ are shown as a function of Sersic index. The blue lines displays the calculation for a Sersic function based on Equation~\ref{eqn:r20r50}. Grey crosses display measurements of isolated galaxies in the GOODS-South field using direct integration of the residual corrected surface brightness profile. We find the observations match the calculation based on the Sersic index very well.}
    \label{fig:sersic_ex}
\end{figure}

Given that the sizes of galaxies at high redshift are comparable to the PSF, one cannot simply measure \rtwo\ and \reig\ directly from the surface brightness profile. Thus we choose to calculate  \rtwo\ and \reig\ from the Sersic profile derived by \texttt{GALFIT}~\citep{Peng2010} . For a single component Sersic profile it is straightforward to convert between \rfiv, \rtwo\ and \reig. The fraction of light contained within a projected radius $r$ is
\begin{equation}
    \frac{L(<r)}{L_{\rm tot}} = \frac{\gamma\left(2n, b_n (r/r_{\rm eff})^{1/n}\right)}{\Gamma(2n)}.
\end{equation}
Here, $\gamma$ is the incomplete gamma function, $\Gamma$ is the complete gamma function, and $b_n$ is the solution to the equation $\Gamma(2n) = 2\gamma(2n,b_n)$, which we approximate as $b_n = 1.9992n - 0.3271$ ~\citep{Capaccioli1989}. Comparing $L(<r_{20})$ to $L(< r_{50})$ we derive the following.
\begin{equation}
    \frac{L(<r_{20})}{L(< r_{50})} = \frac{0.2}{0.5} = \frac{\gamma\left(2n, b_n (r_{20}/r_{50})^{1/n}\right)}{\gamma\left(2n, b_n\right)}
    \label{eqn:r20r50}
\end{equation}
For a given value of $n$, we numerically solve Eqn.~\ref{eqn:r20r50} for the value of $r_{20}/r_{50}$. A similar procedure is used to calculate $r_{80}/r_{50}$. We perform this calculation for a range of Sersic indices with results shown in Fig.~\ref{fig:sersic_ex}. For higher Sersic indices $r_{20}/r_{50}$ decreases, corresponding to the steeper central profile, and $r_{80}/r_{50}$ increases, corresponding to the extended wings at large radius. We present fitting formulas for $r_{20}/r_{50}$ and $r_{80}/r_{50}$ as a function of Sersic index, shown below in Eqn.~\ref{eqn:fit_func}. These fitting functions are accurate to within $5\%$ for $n = 0.25-10$.

\begin{equation}
\begin{split}
    \frac{r_{20}}{r_{50}} (n) &= -0.0008 n^3\, +\, 0.0178 n^2\, -\, 0.1471 n\, +\, 0.6294 \\
    \frac{r_{80}}{r_{50}} (n) &= 0.0012 n^3\, -\, 0.0123 n^2\, +\, 0.5092n\, +\, 1.2646
\end{split}
\label{eqn:fit_func}
\end{equation}

Galaxies, especially those at high redshift, do not necessarily follow a Sersic profile, thus it is important to check whether applying the simple calculation discussed above is broadly applicable.
We tested this by employing the technique used in \citet{Szomoru2010} to correct surface brightness profiles for the effects of the PSF. Galaxies are fit with a single component Sersic profile, which is then convolved with the PSF and subtracted from the observed image to obtain the residual image. The residual image is used to calculate the residual flux profile, which is then added to the (unconvolved) best fit Sersic profile to obtain the corrected profile. \rtwo, \rfiv\ and \reig\ are then calculated by integrating this residual-corrected surface brightness profile.

Figure~\ref{fig:sersic_ex} displays the direct measurements of $r_{80}/r_{50}$ and $r_{20}/r_{50}$ ratios for 127 isolated galaxies in the GOODS-South field. We select these galaxies as being isolated if there is not another sources within $\sim 10$ \rfiv. Their size and magnitude distributions matches those of the overall sample. We use the $H_{160}$ images to directly measure the different radii using the residual corrected surface brightness profile as described above. We find that the direct measurements of $r_{80}/r_{50}$ and $r_{20}/r_{50}$ match the simple calculation based on the Sersic profile well. This is consistent with studies that have shown that high redshift galaxies are generally well fit by a single component Sersic profile~\citep{Szomoru2012}. The scatter of the observed points around the Sersic relation does not correlate with Sersic index, redshift, or galaxy type, but it does increase for galaxies with $m_{\rm F160W} > 23$. Given the success in reproducing $r_{80}/r_{50}$ and $r_{20}/r_{50}$ based on the Sersic index alone, we apply this simple calculation to the rest of our sample with the caveat that the values can be uncertain for individual galaxies.

\begin{figure*}
   \centering
    \includegraphics[width = \textwidth]{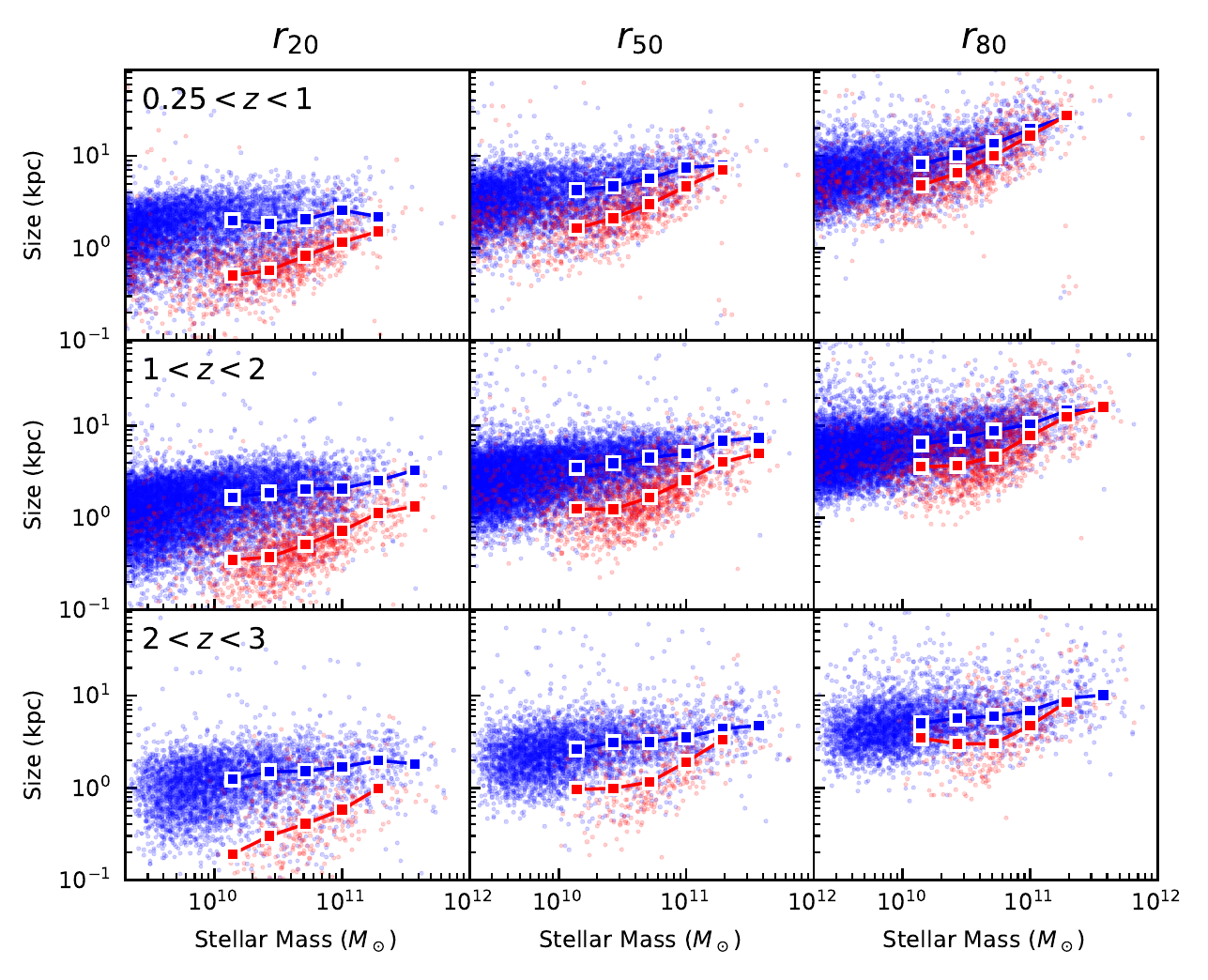}
    \caption{Size-mass distributions of galaxies using different measures of the size. Columns show the size-mass distribution using \rtwo, \rfiv\ and \reig\ as a measure of the size while rows display different redshift bins. Blue and red points show star-forming and quiescent galaxies and the squares show the median sizes in bins of stellar mass. At a given stellar mass the difference between star-forming and quiescent galaxies varies based on which measure of size is used. In \reig\ the two galaxy types largely overlap, where as in \rtwo\ the star-forming and quiescent galaxies follow distinct distributions.}
    \label{fig:mr_all}
\end{figure*}

\section{The Distributions of \rtwo\ and \reig}

\subsection{The Size-Mass Plane}

In Figure~\ref{fig:mr_all} we show the distribution of galaxies in the size-mass plane using three different measures of galaxy size: \rtwo, \rfiv\ and \reig. The size distributions are offset toward larger sizes when going from \rtwo\ to \rfiv\ and \reig, as follows from their definitions. However, we also find that the distributions of star-forming and quiescent galaxies are very different depending on which radius is used. Using \rtwo\ the two populations occupy separate regions of the size-mass plane, with very little overlap. The quiescent galaxies are consistently smaller at a given stellar mass across the entire sample. The \reig-mass plane affords a different view. The star-forming and quiescent populations appear to follow the same distribution, with little difference between the two types of galaxies. The canonical size-mass distribution, using \rfiv, lies between these two extremes. The distribution of galaxies in this plane is often interpreted in the context of the distinct relations that star-forming and quiescent galaxies follow \citep[see][and references therein]{vanderwel2014,Mowla2018}, but as we show in Fig.\ \ref{fig:mr_all} this conclusion depends sensitively on the definition of size. 

\begin{figure*}
    \centering
    \includegraphics[width = 0.49\textwidth]{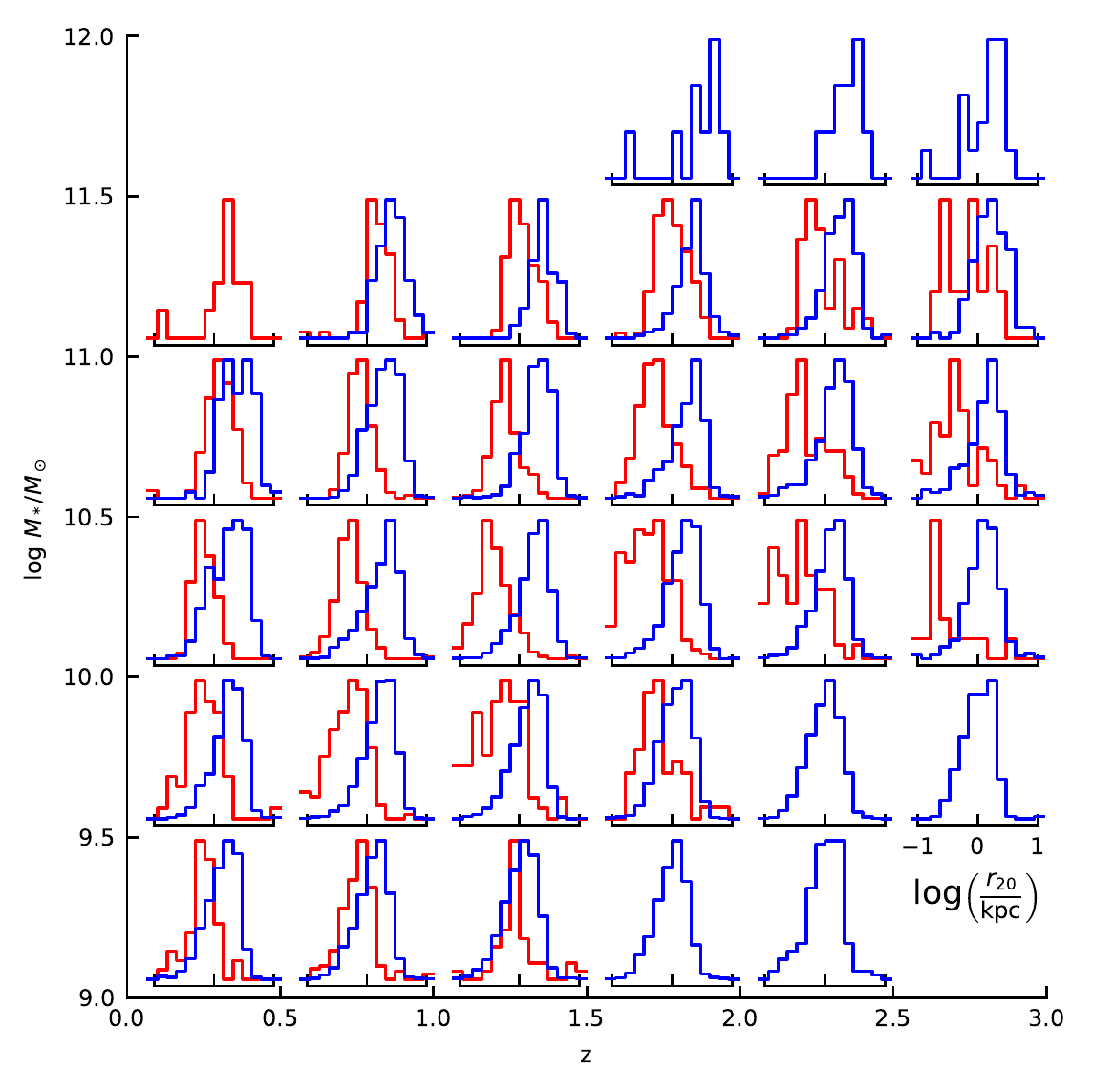}
    \includegraphics[width = 0.49\textwidth]{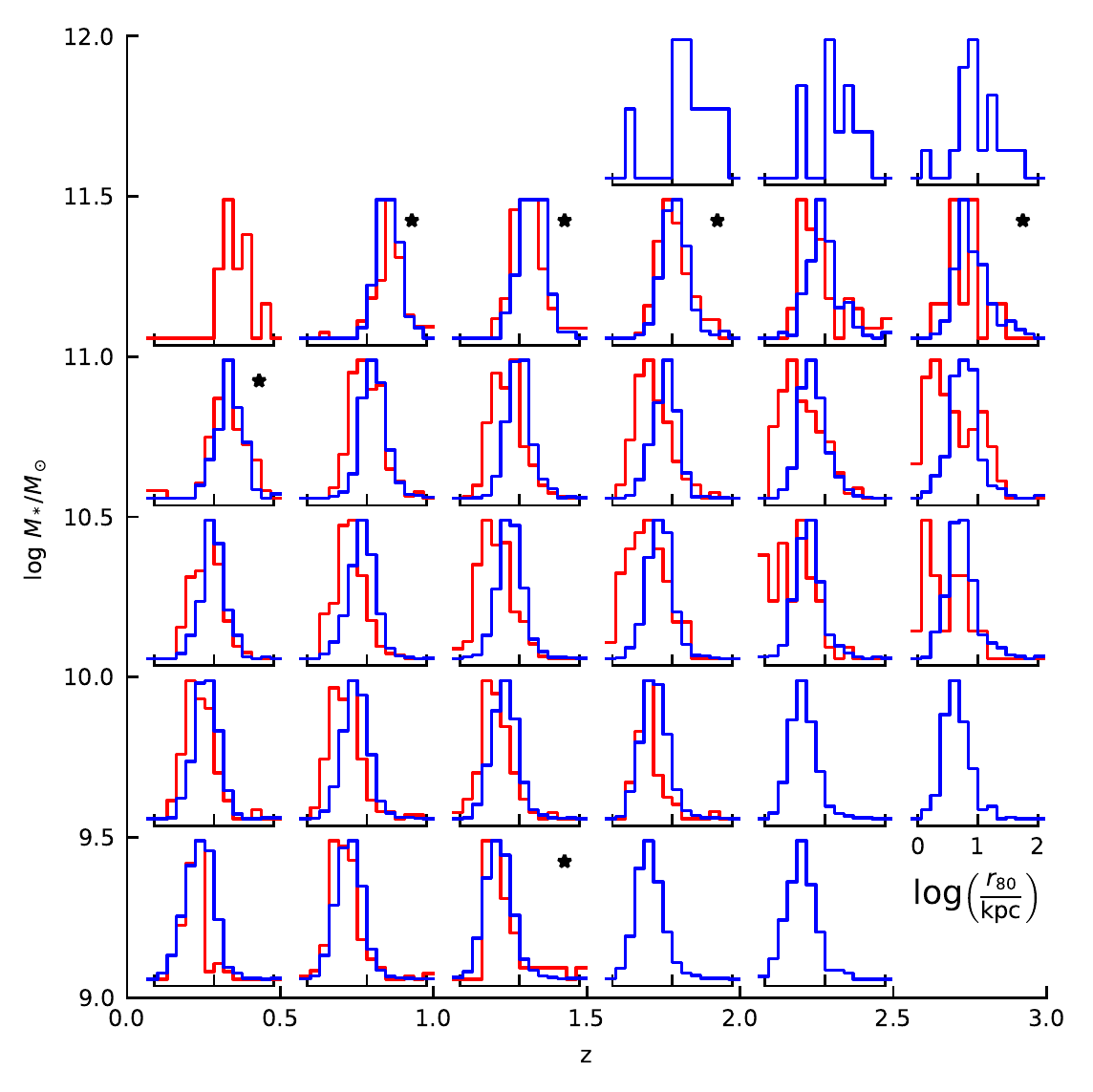}
    \caption{The distribution of galaxies in \rtwo\ and \reig\ for a range of stellar mass and redshift bins. Blue histograms display the distribution of star-forming galaxies and red histograms show quiescent galaxies. We only show histograms where the number of galaxies in that region of parameter space is greater then 8. Additionally, each histogram is normalized to the same height, so the relative heights of the distribution contain no information about the relative number of star-forming or quiescent galaxies in each bin. When considering \rtwo, star-forming and quiescent galaxies follow separate distribution whereas in \reig\ the two types appear to follow the same log-normal distribution. Black stars in the \reig\ panel indicate bins where the Kolmogorov-Smirnov test concludes the distribution of sizes for star-forming and quiescent galaxies could be drawn form the same parent distribution ($p > 0.05$).}
    \label{fig:rdist_bins}
\end{figure*}

The distribution of \rtwo\ and \reig\ for star-forming and quiescent galaxies across a range of stellar masses and redshifts is shown in Fig.~\ref{fig:rdist_bins}. We observe that the two galaxy populations represent two distinct distributions of \rtwo\ while they appear to follow the same distribution in \reig. The bimodality in the distribution of \rtwo\ is most clear for intermediate stellar mass ($ 10< \log M_*/M_\odot < 11 $) and high redshift ($z>1$). Here the peaks of the distributions for star-forming and quiescent galaxies are clearly separated and a valley between the two distributions is apparent. By contrast, the distributions of \reig\ for the two populations are nearly identical. Across the entire range of stellar mass and redshift the peaks and widths of the \reig\ distribution appear at nearly the same location for star-forming and quiescent galaxies. 

\begin{figure*}
    \centering
    \includegraphics[width = \textwidth]{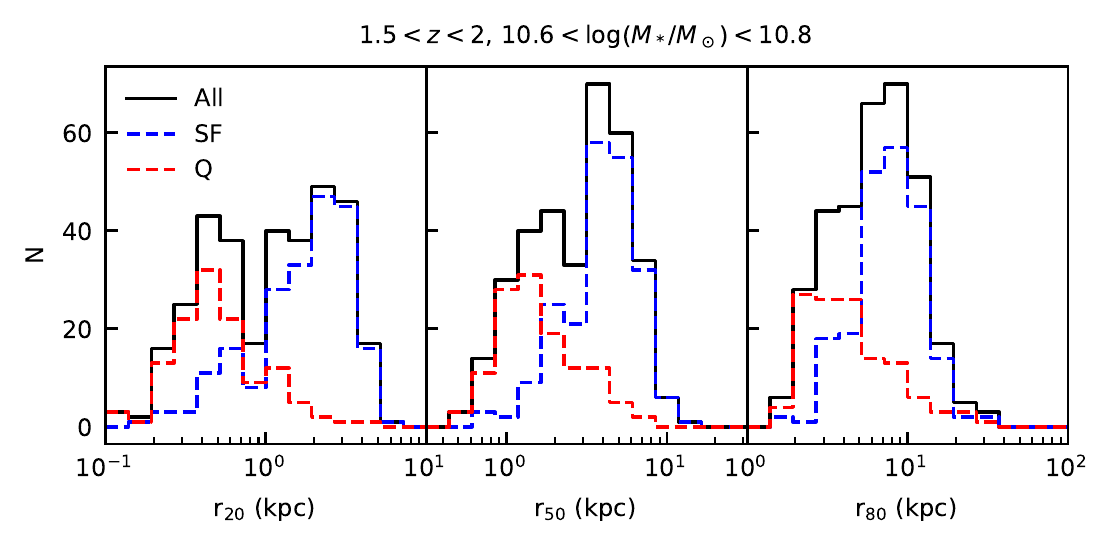}
    \caption{The distribution of \rtwo, \rfiv\ and \reig\ is shown for all galaxies in with $1.5<z<2$ and $10.6 < \log(M_* /M_\odot) < 10.8$. Also displayed are the distributions of star-forming and quiescent galaxies separately. A bimodality in the distribution of galaxies in \rtwo\ can be seen, even without dividing them into star-forming and quiescent. This is confirmed through the Hartigan's dip test. Using Gaussian mixture models, we also show the distribution of \reig\ is consistent with a single log-normal distribution (see text). }
    \label{fig:rdist_sub}
\end{figure*}

\subsection{Bimodality in the Distribution of \rtwo}
To highlight and quantify these trends, we focus on thedistribution of radii in a single stellar mass and redshift bin in Figure~\ref{fig:rdist_sub}. We investigate the overall distribution galaxies, without separating star-forming and quiescent galaxies. The distribution of \rtwo\ appears to be bimodal. To test this hypothesis, we employ Hartigan's dip test~\citep{Hartigan1985}, which tests the null hypothesis that the sample is drawn from a unimodal distribution.\footnote{This test is computed using the R package \texttt{diptest} (https://cran.r-project.org/web/packages/diptest/)} When analyzing the log-space distribution of \rtwo\ in this mass and redshift bin we find $p=0.043$, which means that the null hypothesis of a unimodal distribution can be rejected with $>95$\,\% confidence. As an additional test, we fit one and two component Gaussian mixture models\footnote{We use the \texttt{scikit-learn} python package~\citep{Pedregosa2011}} to the logspace distributions of \rtwo, \rfiv, and \reig\ and compare the Bayesian information criterion (BIC) of each model. Unsurprisingly, the distributions of \rtwo\ and \rfiv\ are better fit by the two component model ($\rm \Delta BIC = BIC_{1\, comp.} - BIC_{2\, comp.} = 49.5\ and\ 41.5$ respectively). Interestingly, for \reig\ we find it is better fit by the single component model ($\rm \Delta BIC = -5.3$).

In the top panel of Figure~\ref{fig:quant} we broaden this analysis and quantify the separation of the distributions of star forming and quiescent galaxies as a function of mass, redshift, and size definition. This is done through the Ashman's $D$ parameter~\citep{Ashman1994}, given by
\begin{equation}
    D = \sqrt{2}\ \frac{ \left| \mu(\log r_{\rm SF}) -  \mu(\log r_{\rm Q}) \right| }{\sqrt{ \sigma (\log\, r_{\rm SF})^2 +  \sigma (\log\, r_{\rm Q})^2} }
    \label{eqn:D}
\end{equation}
Here, $\mu$ is the mean of each galaxy population and $\sigma$ is the standard deviation, which we estimate using the biweight location and scale respectively~\citep{Beers1990}. In the ideal case of a combination of two identical Gaussian distributions the combined distribution shows two distinct peaks if $D>2$~\citep{Everitt1981}. This threshold of $D>2$ is also used more broadly to indicate when a distribution is bimodal, regardless of the functional form. The Ashman D values for \rtwo, \rfiv, and \reig\ are shown in Fig.~\ref{fig:quant} as a function of stellar mass and redshift. At all masses and redshifts the difference between star forming and quiescent galaxies increases when going from \rtwo\ to \rfiv\ and from \rfiv\ to \reig. For \rtwo\ there is significant bimodality with $D>2$ at all stellar masses in the range $1\times 10^{10}< M_*/M_{\odot} < 5\times 10^{10}$.  The Ashman $D$ value decreases at large stellar masses ($M_* > 10^{11} M_\odot$) for all size definitions, echoing the results of \citet{Mowla2018} for \rfiv. The Ashman D value for \reig\ is $\ll 2$ at all stellar masses and redshifts, consistent with the GMM analysis.

\subsection{Implications for the observed scatter in the size-mass relation}

\begin{figure}
    \centering
    \includegraphics[width = \columnwidth]{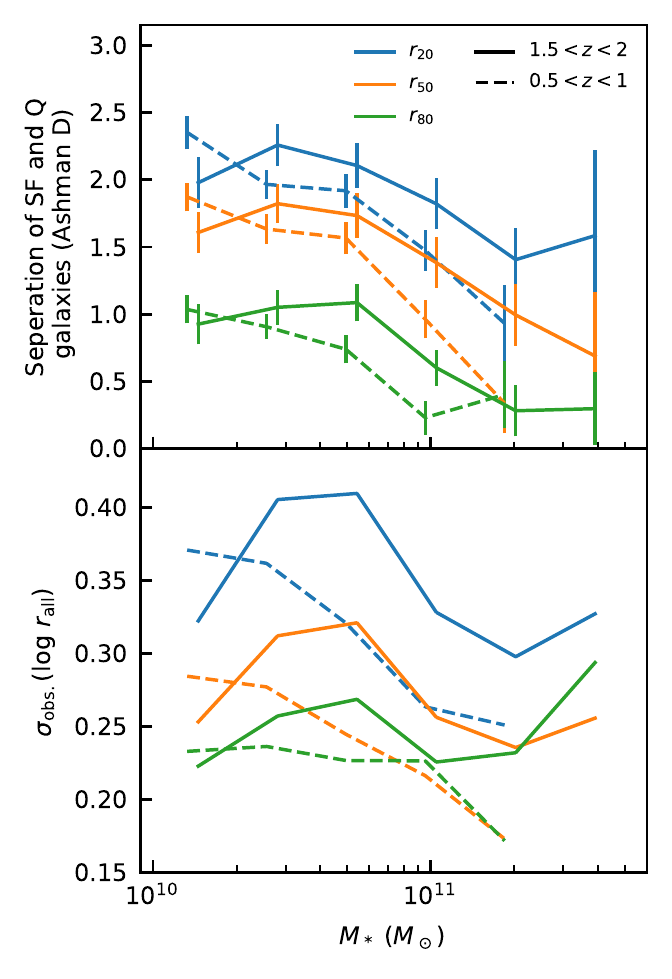}
    \caption{{\em Top panel:} Evidence for bimodality in size distributions as quantified using the Ashman D parameter (Eq.~\ref{eqn:D}). {\em Bottom panel:} The observed scatter, estimated using the bi-weight scale, in \rtwo, \rfiv\ and \reig\ as a function of stellar mass and redshift. The reduced bimodality for \reig~leads to the smaller scatter.}
    \label{fig:quant}
\end{figure}

The fact that the separation of star forming and quiescent galaxies changes for different size definitions has implications for the scatter in the overall size-mass relation: it is significantly smaller for \reig\ than for \rfiv\ and (particularly) \rtwo. This is demonstrated in the bottom panel of Fig.\ \ref{fig:quant}. The observed scatter, estimated using the bi-weight scale, is larger in \rtwo\ than in \rfiv\ by 0.08 dex, due to the fact that the distributions of star-forming and quiescent have a larger separation. The scatter in \reig\ ($\approx 0.25$\,dex, independent of mass and redshift) is generally smaller than in \rfiv. We note that the observed scatter for the quiescent and star-forming galaxies as separate populations is also $\approx0.25$~dex at all masses and redshifts, regardless of the choice of size indicator. This implies that the reduction of the scatter in \reig\ with respect to \rtwo\ and \rfiv\ can be attributed to the fact that the size distributions of star forming and quiescent galaxies overlap in \reig. We are showing the observed scatter in the sizes of galaxies which is the combination of intrinsic scatter and observational uncertainty \footnote{A similar analysis was done in \citet{vanderWel2012} who concluded that the observed scatter in \rfiv\ is dominated by intrinsic scatter in this regime}. To decouple these two quantities would require a careful analysis of the observational procedures and how they affect uncertainties in size measurements. Instead, our goal is to compare the relative scatter of different measures of the size.

\section{Discussion} \label{sec:disc}
In this paper we investigated the size-mass distribution of galaxies if \rtwo\ or \reig, the radii containing 20\% of 80\% of the light, is used instead of the traditional measure of \rfiv. 
When using \rtwo\ we find strong evidence of bimodality in the size distribution at fixed mass; to our knowledge, such a structural bimodality has not been observed before. The two peaks correspond to quiescent galaxies and star forming galaxies.  When using \reig\ the size distribution is narrow and star forming and quiescent galaxies follow very similar size-mass relations at all redshifts.
The results presented here could have been anticipated from the well-known relations between quiescence, mass, size and Sersic index. Specifically, quiescent galaxies are observed to have a higher average Sersic index, which means that $r_{20}/r_{50}$ ($r_{80} / r_{50}$) is lower (higher) when compared to star-forming galaxies. In this sense, the results presented here can be seen as a re-casting of these relations into a convenient form.

Understanding the distribution of light within galaxies aids our understanding of how they assembled~\citep{Hill2017,Huang2018}, and the \rtwo\ and \reig\ distributions may highlight
specific and distinct physical processes.  Based on our results it seems likely that
\rtwo\ is related to processes which affect star formation and quenching. Specifically there appears to be a connection between the structural bimodality discussed in this study and the well known color/ sSFR bimodalities~\citep{Strateva2001, Baldry2004}. It had already been recognized that these bimodalities are connected to the central density of galaxies \citep{Barro2014, vandokkum2015,Whitaker2017,Tacchella2017}. These studies suggest a central density or velocity dispersion threshold above which galaxies quench. At fixed stellar mass, galaxies with a lower \rtwo\ have a higher central density. Therefore, these quenching thresholds are qualitatively consistent with the clean separation of star-forming and quiescent galaxies in \rtwo.

Turning to \reig, this provides a reasonable proxy of the total baryonic extent. At the highest masses typical values of \reig\ reach $\sim 20$\,kpc, and given the similarity of the distributions of star-forming and quiescent galaxies in the \reig-mass plane it is tempting to link this size to halo properties. Several studies have suggested a constant scaling between stellar and halo radius~\citep{Kravtsov2013, Somerville2018}. This connection between \reig\ and the halos of galaxies is explored further in an accompanying paper, Mowla et al.\ (2019). We note that the differences between \rtwo\ and \reig\ can also be interpreted in the context of dynamical time scales; for massive galaxies these are typically a factor of $\sim 20$ longer at \reig\ than at \rtwo. \rtwo\ is therefore sensitive to processes that can change rapidly, such as star formation rates or nuclear activity, whereas \reig\ should be more or less immune to those.

The work presented here is an initial investigation into the differences in the galaxy size-mass distribution when using \rtwo, \rfiv\ and \reig, with more detailed analyses to follow. We have not quantified the evolution of the slope or normalization of the size-mass relation of \rtwo. Describing these trends may give insight into galaxy quenching through cosmic time. For \reig, we refer the reader to Mowla et al. (2019), which details the evolution of the \reig-mass distribution and its connection to halo properties. Another improvement will be measuring the mass profile of galaxies. Recent studies have shown a relatively constant offset between mass-weighted and light-weighted \rfiv\ \citep{Szomoru2013, Mosleh2017}, but it is not clear whether that would also apply to \rtwo\ and \reig. Finally, it is important to continue developing non-parametric techniques for measuring the surface-brightness profiles of high-redshift galaxies. Given current facilities it is a technical challenge to map the inner structure of high redshift galaxies, as the effective radius is comparable to the width of the PSF. Planned AO instruments on 30m class telescopes, which are proposed to provide a factor of $\sim 10$ better resolution, will allow us to map the inner structure of high redshift galaxies directly. 

\acknowledgments

T.B.M. would like to thank Patricia Gruber and the Gruber foundation for their generous support of the work presented here.  AvdW acknowledges funding through the H2020 ERC Consolidator Grant 683184.

%

\vspace{5mm}









\end{document}